\documentclass{aa}
\usepackage{epsf}

\begin{document}
\setcounter{footnote}{1}

\thesaurus{01      
     (08.14.1; 
      13.07.1)}
     
\title{GRB as explosions with standard power}

\author{K.A.~Postnov\inst{1,2},
M.E.~Prokhorov\inst{2}, \and V.M.Lipunov\inst{1,2}}

\institute{
Faculty of Physics, Moscow State University,
\and Sternberg Astronomical Institute, Moscow State University,
                119899 Moscow, Russia
}
\date{Received ... 1999, accepted ..., 1999}
\maketitle

\begin{abstract}

We show that the distribution of observed energies of GRB 
with known redshifts can be 
explained  by the hypothesis of the standard energy release $E_0=5\times
10^{51}$ ergs. Two situations are possible, either the beaming angle 
differs from burst to burst, or there is a universal emssion diagram 
in each burst, the observed difference being due to different viewing angles.


\end{abstract}

\noindent

\section{Introduction}

Standard energy release 
is a general phenomenon in astrophysical sources with (nearly)
fixed masses. The most 
bright example is provided by core collapse supernova explosions.
When a neutron star results from such a collapse,
its binding energy $\Delta E=0.15 M_\odot c^2$ 
is released in the form of neutrino. Measurements of neutron star
masses all indicate a strikingly very narow range 
near 1.4 $M_\odot$ (Thorsett and Chakrabarty 1999),
although the underlying fundamental reason for this is still unknown. 
Another example is type Ia supernovae involving 
explosions of a white dwarf with the Chandrasekhar mass.
Apparently, a similar standard energy must be released 
during binary neutron star coalescences, mostly in the form of
gravitational waves (roughly, 90\%) and neutrino (roughly, 10\%), 
as follows from numerical calculations (Ruffert and Janka 1998).

Can gamma-ray bursts (GRB) join the class? Although their origin 
has not been yet firmly established, the most probable models
include binary neutron star coalescences (as first suggested
by Blinnikov et al. 1984) and collapses of very massive stars (e.g.
Woosley 1993, Paczy\'nski 1998). Note that 
assuming GRB as standard candles and using the position of GRB 970228 (the first 
GRB with low-energy afterglow) on the $\log N\;$--$\;\log S$ diagram,
its redshift was predicted to be $z=0.7\pm 0.1$ immediately after 
its discovery (Lipunov, Postnov, Prokhorov 1997; Lipunov 1998), 
which is in the excellent agreement with measurements 
of the redshift of the host galaxy 
of this GRB ($z=0.695$) made two years later (Djorgovski et al. 1999).
This may be not a pure coincidence.   

Intense optical studies of GRBs have resulted in a rapidly  
increasing number of redshift measurements. Now redshifts of
8 GRBs are known (see Table 1). This enables us to calculate
distances and effective (i.e. assuming spherical symmetry) energy 
release in gamma-rays $E_\gamma$. This energy, as seen from Table 1, 
varies in a broad range from $\sim 5\times 10^{51}$~erg to $\sim 2\times
10^{54}$~erg. This fact can be treated as a wide proper
luminosity function of GRBs. 

The actual energy release $E_0$ may be significantly lower than
$E_\gamma$ due to a possible beaming of gamma-ray emission. If beaming angle
$\theta \ll 1$, then the actual energy release 
$E_0\simeq E_\gamma\frac{\theta^2}{4}$.

In the general case, the distribution of $E_\gamma$ may be affected by 
both the proper energy release distribution $f(E_0)$ and the beaming
factor distribution $f(\theta)$. Here we show that
the existing observations of GRBs conform with the hypothesis of
a standard energy release $E_0$ in the underlying GRB explosions, and 
the apparent dispersion in the detected energy can be explained by 
the beaming factor $E_0/E_\gamma$. 
The observed detected energy can be explained by both 
distribution of beaming angles and by some universal shape of 
the emission diagram in GRBs.

\section{Three groups of the observed GRB energies}

In Table 1 we list eight GRBs with measured redshifts. 
We excluded GRB 980425, which possibly relates to SN 1998 bw in 
a close (40 Mpc) galaxy. 
We included into the Table 1 GRB 980329 with 
high redshift $\sim$5, which was indirectly 
deduced (Palazzi et al.
1998; Fruchter 1999) and can be excluded from our consideration. This 
however has an insignificant effect on the final conclusions.

\begin{table}
\caption{}
\begin{tabular}{ccclc}
\hline\hline
GRB & $z$ & $\Delta E_{\rm obs}^{a)}$ & $E_\gamma/E_0$ & Beaming \\
\hline
970228 & 0.695$^{b)}$ & $5.2\times10^{51}$ & $\sim$1 & NO    \\
970508 & 0.835 & $5.3\times10^{51}$ & $\sim$1 & NO    \\
971214 & 3.418 & $2.5\times10^{53}$ & $\sim$50 &      \\
980329 &$\sim$5?&$\sim$$2\times10^{54}$ & $\sim$500& \\
980613 & 1.096 &   $5\times10^{51}$ & $\sim$1       & \\
980703 & 0.966 &   $9\times10^{52}$ & $\sim$20      & \\
990123 & 1.60  & $1.6\times10^{54}$ & $\sim$300&YES  \\
990510 & 1.62$^{c)}$  & $1.4\times10^{53}$ & $\sim$30& YES  \\
\hline\hline
\end{tabular}\\
a) Data from Briggs et al. 1999\\
b) Djorgovski et al. 1999\\
c) Vreeswijk et al. 1999\\
\end{table}

Of these eight GRBs, 
three 
(GRB\,970228, GRB\,970508, GRB\,980613) display about the same effective
energy release of $(4.2$--$5.3)\times 10^{51}$ ergs. They form the 
weakest group of GRBs. Another three GRBs
(GRB\,971214, GRB\,980703, GRB\,990510)
form the intermediate group with the effective energy release
$9\times 10^{52}\,$--$\,2.5\times 10^{53}$~ergs,
i.e. 20--50 times as bright as the first group.
The last two GRBs (GRB~980319 and GRB~990123) are the brightest among all
GRBs with the effective energy release $2.4\times 10^{54}$ and $1.6\times
10^{54}$~ergs, respectively (500 and 300 times brighter than the first
group). We should note that the dispersion in energy release for GRBs 
from the first group is much smaller than for bursts from the second and the
third groups. The mean reshift in these groups shows tendency to grow 
with energy: $\langle z_1\rangle = 0.875$,  $\langle z_2\rangle = 2.0$, 
 $\langle z_3\rangle = 3.3$. This correlation is natural in the hypothesis 
of beaming: the narrowest beams corresponding to the largest observed
energies should be detected less frequently from a given 
radius and thus will be observed from
larger distances to give a comparable number of detections with
weaker energies.

GRB~990123 shows some evidence for a 
significant beaming $\theta\sim 0.1$ (Kulkarni et al. 1999)
which is deduced from the break in the optical afterglow light curve.
Such a break is expected to occur when the Lorentz-factor
of the expanding relativistic shell $\Gamma$ matches the inverse 
beaming angle $\Gamma\sim 1/\theta$ (Rhoads 1999). The true energy release 
for this GRB thus becomes $E_0\sim 4\times 10^{51}$~erg, close to the 
weakest group of GRB effective energy. This may imply that GRBs from the weakest
group occur almost spherically-symmetrically and $E_0=E_\gamma$ for this
bursts and the energy release $E_0\simeq 5\times 10^{51}$~ergs is a 
fundamental value for {\it all} GRBs.
A smooth
broad-band change is observed in the slope of the afterglow of the recent
GRB~990510 (Harrison et al. 1999), which can also be explained by beaming, so the 
actual energy release in this case is again smaller. In contrast, no
indication
of beaming is seen for longest observed optical afterglows of 
GRB~970508 and GRB~970228, and for them the observed energy release
$E_\gamma$ is approximately the same, about $5\times 10^{51}$ erg.

\begin{figure}
\epsfxsize=\hsize
\epsfbox{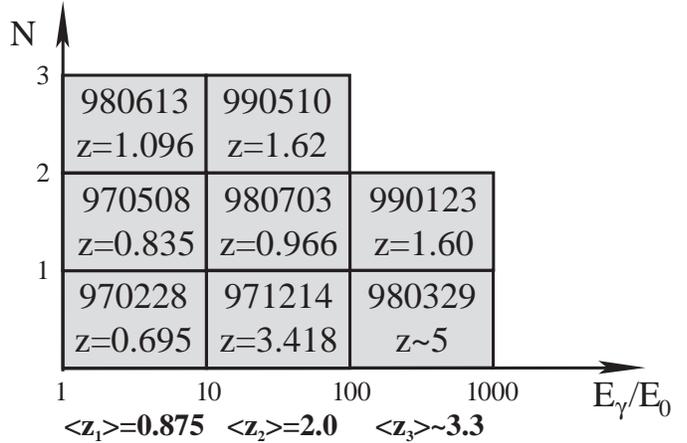}
\caption{The distribution of GRBs with known redshifts 
by their $E_\gamma/E_0$ ratio. The individual redshift of GRBs
and the  mean redshift in each group are indicated}
\end{figure}

\section{Standard GRB energy release}

Let us postulate the standard energy release in all GRBs to be $E_0=5\times
10^{51}$ ergs. 
The three groups of GRBs by their $E_\gamma/E_0$ ratio 
are schematically shown in Fig. 1. This distribution   
is fairly flat and can be treated in two different ways. It 
can both reflect the dispersion of 
in beaming factors of individual bursts  and  be explained 
by a universal shape of the gamma-ray emission diagram of GRBs
with the standard energy release. 

\subsection{Beaming angle distribution}

Consider first beaming angle distribution. In this hypothesis 
the standard energy $E_0=5\times 10^{51}$ ergs is assumed to be deposited
into cones with different opening angles $\theta$. 
 For the sake of simplicity we
shall assume Euclidean space. Then the observed energy in each group is
\begin{equation}
E_{\gamma, i}=\frac{E_0}{(\Omega_i/4\pi)}, \quad i=1,2,3\,,
\end{equation}
and by assumption 
\begin{equation}
\frac{\Omega_1}{4\pi}=1, \quad \frac{\Omega_2}{4\pi}=\frac{1}{30}, \quad
\frac{\Omega_3}{4\pi}=\frac{1}{300}\,.
\end{equation} 
Let $n_0$ be the spatial concentration of GRB sources, 
$n_i$ be the fraction of sources with corresponding $\Omega_i$ so that 
\begin{equation}
n_1+n_2+n_3=1\,.
\end{equation}
The limiting distance from which a GRB form i-th group can be observed
is $R_i=R_1(E_{\gamma,i}/E_0)^{1/2}$ so that the number of i-th events
potentially observed is 
\begin{equation}
N_i=n_0n_i\left(\frac{\Omega_i}{4\pi}\right) \frac{4\pi}{3} R_1^3
\left(\frac{E_{\gamma,i}}{E_0}\right)^{3/2} 
\end{equation}   
Solving Eqs. (1-4) we obtain $n_1=82\%$, $n_2=15\%$, $n_3=3\%$
and the corresponding opening angles $\theta_2=15^o$, $\theta_3=3^o$
(we used $\theta_i\approx\sqrt{\Omega_i/2\pi}$ which is valid for 
$\Omega_i\ll 4\pi$). This distribution is illustrated by Fig. 2.

\begin{figure}
\epsfysize=\columnwidth
\centerline{\epsfbox{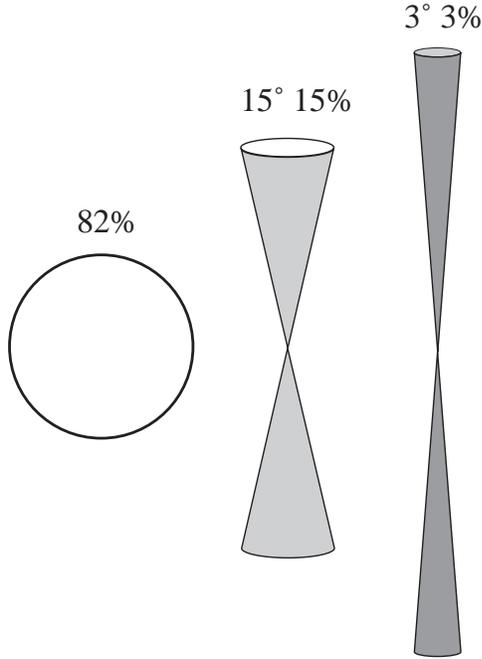}}
\caption{Distribution of beaming angles of GRBs assuming 
standard energy release $E_0=5\times 10^{51}$ ergs derived from the
observed statistics. Figures show the relative number of events with 
opening angles $\theta$}.
\end{figure}

\subsection{A universal diagram of gamma-ray emission}

In this hypothesis we assume that for some reason in {\it all}
GRB the standard energy $E_0$ is always deposited into a universal
emission diagram, with  
the observed number 
of events from different GRB groups coming  from different viewing angles with 
respect to the symmetry axis of the diagram (Fig. 3). Then Eq. (1) modifies
into
\begin{equation}
E_{\gamma, i}=\frac{\epsilon_iE_0}{(\Omega_i/4\pi)}, \quad i=1,2,3\,,
\end{equation}
where $\epsilon_i$ characterizes the fraction of the total energy which
is collimated into the cone $\omega_i$, Eq. (2) reads
\begin{equation}
\frac{\Omega_1}{4\pi}=1, 
\quad \frac{\Omega_2}{4\pi}=\frac{1}{30}\frac{\epsilon_2}{\epsilon_1}, \quad
\frac{\Omega_3}{4\pi}=\frac{1}{300}\frac{\epsilon_3}{\epsilon_1}\,.
\end{equation} 
normalization (3) transits into
\begin{equation}
\epsilon_1+\epsilon_2+\epsilon_3=1
\end{equation} 
(here we neglect overlapping between the cones since $\Omega_1\gg\Omega_2\gg
\Omega_3$; this would yield a minor correction),
and Eq. (4) becomes
\begin{equation}
N_i=n_0\left(\frac{\Omega_i}{4\pi}\right) \frac{4\pi}{3} R_1^3
\left(\frac{\epsilon_i E_{\gamma,i}}{E_0}\right)^{3/2}. 
\end{equation}  

Solving system (6-8) we arrive at $\epsilon_1=74\%$, $\epsilon_2=21\%$, 
$\epsilon_3=5\%$ collimated into the cones with opening angles 
$\theta_1=90^o$ (isotropic emission), 
$\theta_2=20^o$, and $\theta_3=3^o$, respectively.
Note that in this variant the total energy release turns out
to be by $\sim$22\% larger and is $\sim 6\times 10^{51}$ ergs.
This  diagram is illustrated by Fig. 3. 

\begin{figure}
\epsfysize=\columnwidth
\centerline{\epsfbox{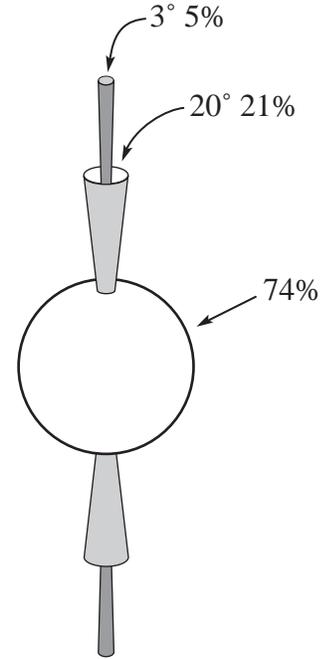}}
\caption{The assumed universal emission diagram of GRBs
can be roughly divided into three parts: a quasi spherically symmetric
part, into which 74\% of the total energy is emitted, and two more narrow
cones with opening angles of 20 and 3 degrees, into which 21\% and 5\%
of the total energy is collimated, respectively. The total energy release 
is $6\times 10^{51}$ ergs.}
\end{figure}

\section{Conclusions}

We have shown that the observed distribution of GRB energies can be 
explained  by the hypothesis of the standard energy release $E_0=5\times
10^{51}$ ergs. Two situations are possible, either the beaming angle 
differs from burst to burst, or there is a universal emssion diagram 
in each burst, the observed difference coming from different viewing angles.
At this stage, we don't discuss any physical model for such a diagram.
Possibly, GRBs observed as different parts of thus shaped energy release
can have different physical features (for example, some spectral or
temporal peculiarities). 

The opposite point of view is that the energy release in GRBs can vary by 
orders of magnitude and beaming is not significant (e.g. Totani 1999). 
Although so far we can not distinguish between different possibilities (i.e.
whether the observed $E_\gamma$ distribution is due to different 
beaming, or due to different true energy release, or both),
the increased statistics of GRB redshift measurements, which is expected 
in the near future, can be used to discriminate between these points of
view. 

The increase in statistics, however, cannot discriminate between the
two possible variants discussed in this paper, because both beaming angle
distribution and the universal diagram
can discribe an arbitrary number of groups of GRBs with an arbitrary number 
of events inside each group. The hypothesis of a standard energy release 
can be discarded if  (a) a GRB with mush smaller $E_\gamma$ than $E_0$
is observed; (b) an inconsistency is found of the predicted cone angle $\theta$ 
to the value mesaured by some means; (c) the mean redshift of GRB in groups 
does not increase with observed energy $E_\gamma$.  Apart form a doubtful
case of GRB~980425 possibly associated with SN 1998bw, no GRBs with smaller
than $10^{51}$ erg energy have been observed. The determination of beaming
angles from the existing observational data is also not very accurate now
(see Sari, Piran and Halpern 1999 for more detail). The mean redshift of
GRBs do increase, on average, with observed energy (see Fig. 1).

The authors thank S.I.Blinnikov, N.I.Shakura and I.G.Mitrofanov  
for discussions. 
The work is partially supported by
Russian Fund for Basic Research through Grant 99-02-16205
and INTAS grant 96-315.


\begin{thebibliography}{}

\bibitem{}Blinnikov S.I., Novikov I.D., Perevodchikova T.V., 
Polnarev A.G., 1984, SvA Lett. 10, 177

\bibitem{}Briggs M.S., Band D.L., Kippen R.M., et al., 1999, Ap in press
(astro-ph/9903247)
\bibitem{}Djorgovski S.G., Kulkarni S.R., Bloom J.S., 
Frail D.A., 1999, GCN 289

\bibitem{}Fruchter A. et al. 1999, ApJ in press

\bibitem{}Harrison F.A., et al., 1999, astro-ph/9905306

\bibitem{}Kulkarni S.R., et al., 1999, Nature 398, 389
 
\bibitem{}Lipunov V.M., 1998, in Highlights of Astronomy 11B, 779

\bibitem{}Lipunov V.M., Postnov K.A., Prokhorov M.E., 1997, astro-ph/9703181

\bibitem{}Paczy\'nski B., 1997, ApJ 499, L45

\bibitem{}Palazzi E., et al. 1998, A\&A 336, L95

\bibitem{}Rhoads J.E., 1999, ApJ submitted, astro-ph/9903399

\bibitem{}Ruffert M., Janka H.-Th., 1998, A\&A 338, 535

\bibitem{}Sari R., Piran T., Halpern J., 1999, astro-ph/9903339

\bibitem{}Thorsett S.E., Chakrabarty D., 1999, ApJ 512, 288

\bibitem{}Totani T., 1999, MNRAS submitted, astro-ph/9907001

\bibitem{}Vreeswijk P.M., Galama T.J., Rol E., et al., 1999, GCN 324 


\bibitem{}Woosley S.E., 1993, ApJ 405, 273 

\end{thebibliography}
\end{document}